\documentclass[1p]{elsarticle}

\usepackage{lineno,hyperref}
\usepackage{amsmath}
\usepackage{ascmac}
\usepackage{here}
\usepackage{color} 
\usepackage{ulem}

\newcommand{\add}[1]{#1}	
\newcommand{\erase}[1]{\if0{#1}\fi}	

\modulolinenumbers[5]

\journal{Journal of \LaTeX\ Templates}

\begin{document}

\begin{frontmatter}

\title{
Comparison of gap-based and flow-based control strategies using a new controlled stochastic cellular automaton model for traffic flow
}

\author{Kayo Kinjo}
\address{Graduate School of Science and Engineering, Saitama University, 255 Shimo-Okubo, Sakura-ku, Saitama 338-8570, Japan}
\ead{kayo.kinjo1@gmail.com}

\author{Akiyasu Tomoeda}
\address{Faculty of Informatics, Kansai University, 2-1-1 Ryozenji-cho, Takatsuki-shi, Osaka 569-1095, Japan}
\begin{abstract}


Autonomous vehicles are essential to future transportation systems, potentially reducing traffic congestion.
This study examines the impact of different vehicle control strategies on traffic flow through simulations. We
propose a novel stochastic cellular automaton model, the controlled stochastic optimal velocity (CSOV)
model, which incorporates vehicle control effects. Within the CSOV model, two control strategies are
implemented: gap-based control (GC), which adjusts vehicle velocity to balance the gaps between adjacent
vehicles, and flow-based control (FC), which aims to maintain a consistent local flow between the front and
rear vehicles. Results show that both control strategies improve traffic flow. However, under weaker control,
the GC sometimes resulted in lower flow compared to no control. In contrast, the FC consistently enhanced
flow across control strengths, yielding more robust outcomes. Furthermore, when both strategies achieved
comparable flow rates, the FC provided a more stable velocity distribution under varying traffic densities
than the GC.

\end{abstract}


\end{frontmatter}

%
\section{Introduction} \label{sec:intro}
%

Modern life is heavily dependent on automobile-based transportation. 
This automobile-centered transportation system is subject to traffic congestion, which has various adverse effects on our living environment, such as reduced transportation efficiency 
and an increased environmental burden due to exhaust emissions. 
For example, \cite{Schrank2015} reported that in 2014, 
urban Americans drove 6.9 billion extra hours due to traffic congestion, consumed 3.1 billion gallons of oil, and suffered economic losses of 160 billion U.S. dollars due to the additional time and oil consumed. 
Hence, easing traffic congestion has become an imperative mission for modern society.

The phenomenon of traffic flow, including traffic congestion, has attracted considerable research interest not only in the fields of traffic and transportation engineering but also in the natural sciences \cite{chowdhury2000, helbing2001, nagatani2002, nagel2003, schadschneider2010, treiber2013};
This is because traffic flow is not only a subject of practical importance for easing traffic congestion 
but also because it can be considered a typical self-driven many-particle system that is far from equilibrium.
Research on traffic flow has a long history, dating back to the first half of the 20th century. 
For example, Greenshield \cite{Greenshield1935} conducted an experimental study \add{where they measured actual traffic} \erase{on traffic flow by measuring actual}flows; he suggested a linear relation between traffic density and velocity, which is still sometimes used for analytical investigations. 
Starting with this pioneering study, research on traffic flow has developed significantly around sophisticated mathematical models, such as 
hydrodynamic models 
(e.g., see \cite{Jafaripournimchahi2022} for a brief overview on recent relevant studies),  
car-following models (e.g., see \cite{wang2023} for a review on existing studies on car-following models),
and cellular automaton models (e.g., see \cite{tian2021} for a review on CA models).
In the last decade or so, as the performance and convenience of measurement equipment have improved and accurate traffic data have become more readily available, experimental studies have been vigorously conducted using 
a circular road in homogeneous lane conditions on flat ground \cite{sugiyama2008, nakayama2009, tadaki2013, stern2018}, 
an open road section in a suburban area \cite{jiang2014, jiang2015, jin2015}, 
and the runway and taxiway of an old airport \cite{jiang2018}. 
These significant contributions to traffic flow research have deepened our understanding of the dynamics of traffic flow.

One of the key contributions of these traffic flow studies was the elucidation of the mechanism of \textit{phantom traffic jams}, which occur spontaneously without any apparent reasons, such as bottlenecks, merging, and lane changes. Phantom traffic jams were clarified to be the collective effect of a many-particle system that induces destabilization of the free flow state due to enhancement of fluctuations; when the average vehicle density exceeds a certain critical value, the transition to a jamming state occurs spontaneously \cite{sugiyama2008, nakayama2009, tadaki2013}.
One way to ease traffic congestion is to change the driving behavior of each driver as the origin of phantom traffic jams lies within the collective motion of vehicles. Driving strategies such as \textit{jam-absorption driving} have also been proposed and developed \cite{nishi2013, taniguchi2015, nishi2020, nishi2022}.

Meanwhile, the traffic flow landscape has begun to change dramatically in recent years with the advent of autonomous vehicles. 
Autonomous vehicles are expected to lead the next paradigm shift in the field of transportation, and the widespread use of autonomous vehicles is highly anticipated to alleviate traffic congestion.
However, the characteristics of mixed traffic between human-driven vehicles (HVs) and autonomous vehicles (AVs) have to be investigated and understood because realization of a fully-automated driving society will take time and mixed traffic will be a normalcy in the near future.
Numerous studies have been conducted on mathematical models of autonomous vehicles and mixed traffic flow with HVs \cite{cui2017, zhu2018, yao2019, yamamoto2018,  yu2021}
(e.g., see \cite{yu2021} for a review of mathematical models for autonomous vehicles).

As the strategies of autonomous vehicles target the movements of individual vehicles, they are more suitable for microscopic models, such as car-following and cellular automaton models, where the focus is explicitly on each individual. 
Examples of previous studies built on the car-following model are as follows: Cui et al. \cite{cui2017} presented a theoretical analysis on the potential of a single connected and autonomous vehicles (CAV) in a ring-road scenario; Zhu et al. \cite{zhu2018} proposed a new car-following model with adjustable sensitivity and smooth factor to describe the autonomous vehicle's movement to balance the front and back headway and analyzed their impacts on traffic flow; Yao et al. \cite{yao2019} studied the stability of mixed traffic flows comprising CAVs and HVs and showed that under certain conditions, mixed traffic flows would not be affected by the velocity for a certain penetration rate of the CAVs and a desired time headway.
Conversely, Ye and Yamamoto \cite{yamamoto2018} proposed an extended CA model that incorporates CAVs in the heterogeneous traffic flow and numerically investigated the possible impact of CAVs on road capacity under different penetration rates.

Although research on mathematical models for autonomous vehicles is increasingly active, most models describe the control based on velocity or distance; none have considered the flow rate or incorporated it explicitly into the model.
The essence of congestion mitigation lies in increasing the number of vehicles passing through a measurement point per unit of time and maximizing the flow rate. 
As this flow rate can be obtained by the product of velocity and density, the strength of focusing on flow rate is to maximize flow rate by balancing both velocity and distance between vehicles. For instance, even if a series of vehicles keep uniform spacing, if they are driven so slowly that the flow rate is lowered, we cannot say this is a solution to traffic congestion.

\begin{figure}[ht]
    \centering
    \includegraphics[width=0.85\linewidth]{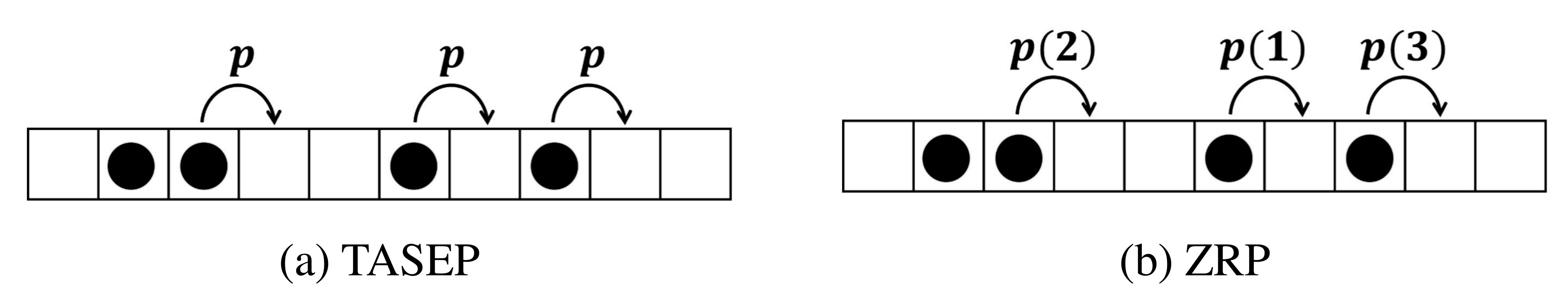}
    \caption{Schematic view of TASEP and ZRP under a periodic boundary condition.}
    \label{fig:tasepzrp}
\end{figure}

The car-following model has an advantage in theoretical analyses such as stability analysis\add{. However,}\erase{, but as}  described in \cite{tian2021}, 
the cellular automaton model also has advantages, such as computational efficiency, simplicity of control algorithms, and flexible scenario reproducibility. 
These advantages are effective and important factors in large-scale simulations.
Therefore, this study proposes a new model based on the stochastic optimal velocity (SOV) model \cite{kanai2005} that encompasses the totally asymmetric simple exclusion process (TASEP) \cite{ASEP} and the zero range process (ZRP) \cite{ZRP}. 
\add{In the modeling of traffic flow, TASEP and ZRP can be interpreted in terms of their dynamics as follows, respectively \cite{schadschneider2010}. 
In TASEP (Figure \ref{fig:tasepzrp}(a)), particles are allowed to move in only one direction, and if the site directly in front of a particle is empty, the particle moves forward with a certain probability $p$. TASEP is a very simple particle-hopping dynamics and is the paradigmatic model for various traffic models.
On the other hand, ZRP (Figure \ref{fig:tasepzrp}(b)), when interpreted in terms of the corresponding exclusion process, is a model where the hopping probability $p(h)$ varies with the distance $h$ from the particle in front, for particles moving in only one direction, as in TASEP. That is, if $p(h)=const.$ then ZRP is reduced to TASEP.
}
The model can obtain analytically important exact solutions and reproduce metastability, which is an essential feature of traffic flows. 
While maintaining the framework of the stochastic CA model, we propose a new model that incorporates the effect of control. Moreover, we clarify the characteristics of traffic flows with a mixture of controlled vehicles.

The remainder of this paper is organized as follows. First, we
briefly review the SOV model and propose a new controlled SOV
(CSOV) model with two control strategies: gap-based control (GC)
and flow-based control (FC). Next, we present the numerical results
for the CSOV model. Finally, we provide concluding discussions.

\section{Controlled SOV (CSOV) model}\label{sec:csovmodel}

We propose a novel model of controlled vehicles called the CSOV model to explore the difference between vehicle control strategies and their impact on mixed traffic through simulations. The model builds upon the SOV model \cite{kanai2005}, which describes the movement of HVs as a hopping probability of moving to the next cell. 
As most vehicles in use today have a vehicle control system as a driving support rather than an autonomous driving system, this study is modeled based on the concept of a HV with control strategies.
Thus, the CSOV model proposed in this study incorporates an additional term for vehicle control in the SOV model.

\subsection{HVs in the SOV model} \label{subsec:sov}
\begin{figure}[H]
    \centering
        \includegraphics[width=0.9\linewidth]{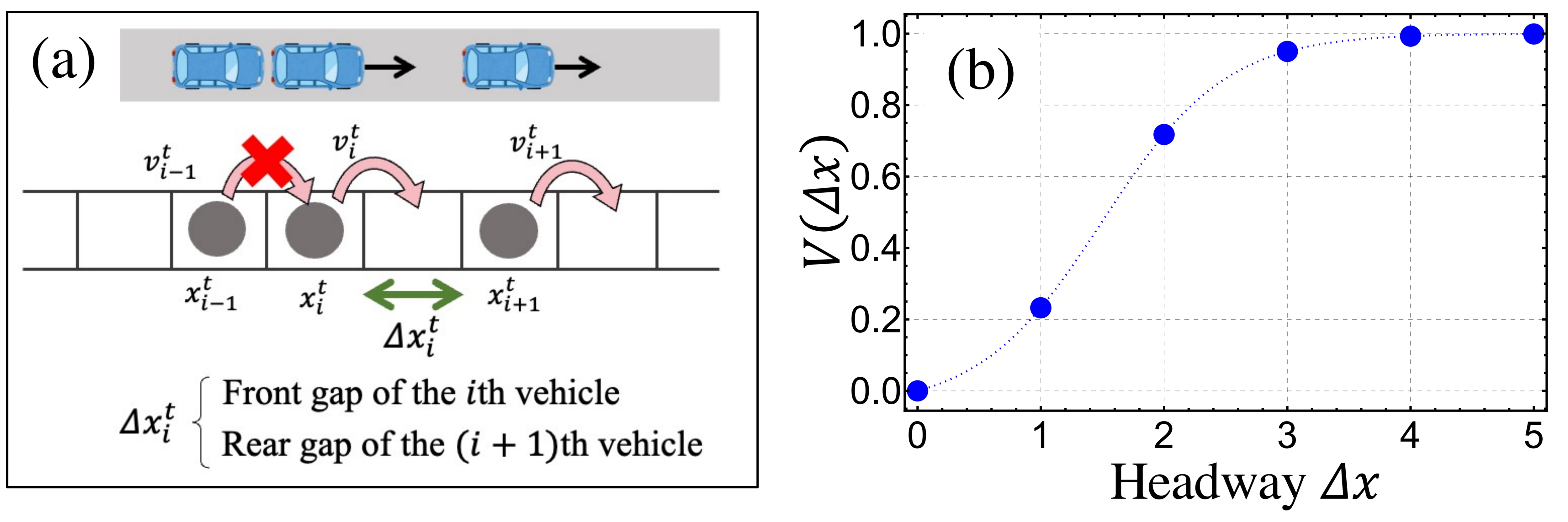}
    \caption{(a) Schematic illustration of the SOV model and (b) OV function. As the headway is discretized in the SOV model, the values of the OV function are represented by the blue points.}
    \label{fig:OVfunction}
\end{figure}

Let us start by briefly introducing the SOV model.
The SOV model is a stochastic cellular automaton model; hence, both time and position are discretized: the road is divided into $L$ identical cells such that each cell can accommodate at most one vehicle at one time step, and the velocity of a vehicle is represented as a hopping probability of moving to the next cell. Figure \ref{fig:OVfunction} (a) shows the schematic illustration of the SOV model.
When the $i$th vehicle is located at position $x_i^t\in\{1,2,3,\cdots,L\}$ at time step $t$, and its velocity is $v_i^{t}\in[0,1]$, the velocity of the $i$th vehicle at the next time step $t+1$  is given by
\begin{align}
    v_i^{t+1} &= (1-\alpha)v_i^t + \alpha V(\Delta x^t_i).
    \label{eq: sov}
\end{align}
The above expression can be regarded as the weighted average between the current velocity $v_i^t$ and an ideal velocity for a given front gap $\Delta x^t_i := x_{i+1}^t - x_i^t -1$, which is given by an optimal velocity (OV) function, in a proportion of $(1-\alpha):\alpha$.
Here, \erase{as in the previous study \cite{kanai2005},}we employ the OV function $V(\Delta x)\in[0,1]$, which is given by
	\begin{align}
		V(\Delta x) &:= \dfrac{\tanh(\Delta x-c)+\tanh(c)}{1+\tanh(c)},
  \label{eq:ovfunction}
	\end{align}
where the constant $c$ was set to \erase{1.0}\add{1.5} throughout this paper\add{, consistent with the value used in the previous study \cite{kanai2005}}. 
Figure \ref{fig:OVfunction}(b) plots the OV function of Equation \eqref{eq:ovfunction}.

\begin{figure}[ht]
    \centering
    \includegraphics[width=0.85\linewidth]{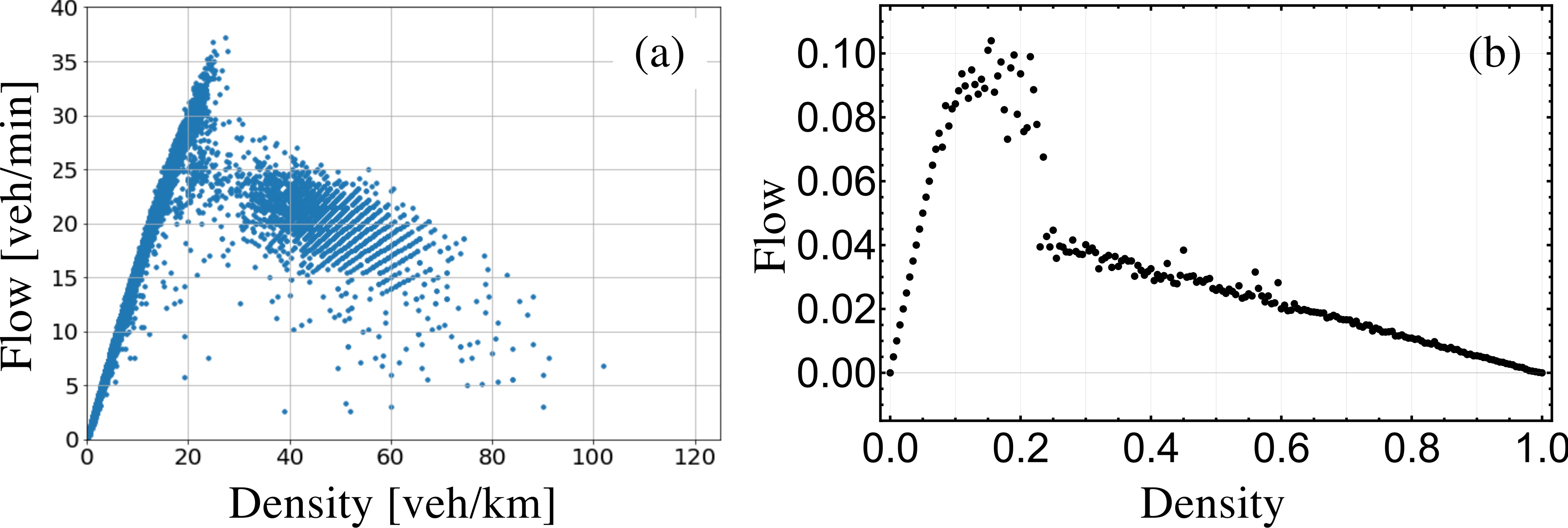}
    \caption{Fundamental diagrams of (a) measurement data and (b) SOV model for a system length $L$ of 200, $\alpha=0.01$, under the periodic boundary condition.}
    \label{fig:FD_comp}
\end{figure}

The SOV model has notable features to describe the HVs.
First, it can reproduce the metastable state commonly found in practical traffic, such as highway traffic. In the metastable state, a subtle perturbation of flow is sufficient to trigger the transition into the congested state.
Figures \ref{fig:FD_comp} (a) and (b) show a fundamental diagram of the measured traffic flow on a highway and that of the SOV model, respectively. Within the low-density region, the flow increases as the density increases (Figure \ref{fig:FD_comp}(a)). However, beyond a certain density, called the critical density, a further increase in the density leads to a discontinuous reduction of the stationary flow. It is also reported that this characteristic is also observed in the SOV model (Figure \ref{fig:FD_comp}(b)).
Moreover, the SOV model includes two solvable stochastic processes: \erase{the} TASEP \cite{ASEP} and \erase{the} ZRP \cite{ZRP}. When $\alpha = 0$ in Equation \eqref{eq: sov}, the SOV model reduces to \erase{the} TASEP, where the velocity $v_i$ remains constant as $v_i=v_i^0$ for $i=1,\cdots, N$. Conversely, when $\alpha = 1$, the velocity depends only on the second term of Equation \eqref{eq: sov}, which represents a function in terms of the front gap. This configuration corresponds to \erase{the} ZRP.

\begin{figure}[H]
    \centering
        \includegraphics[width=0.4\linewidth]{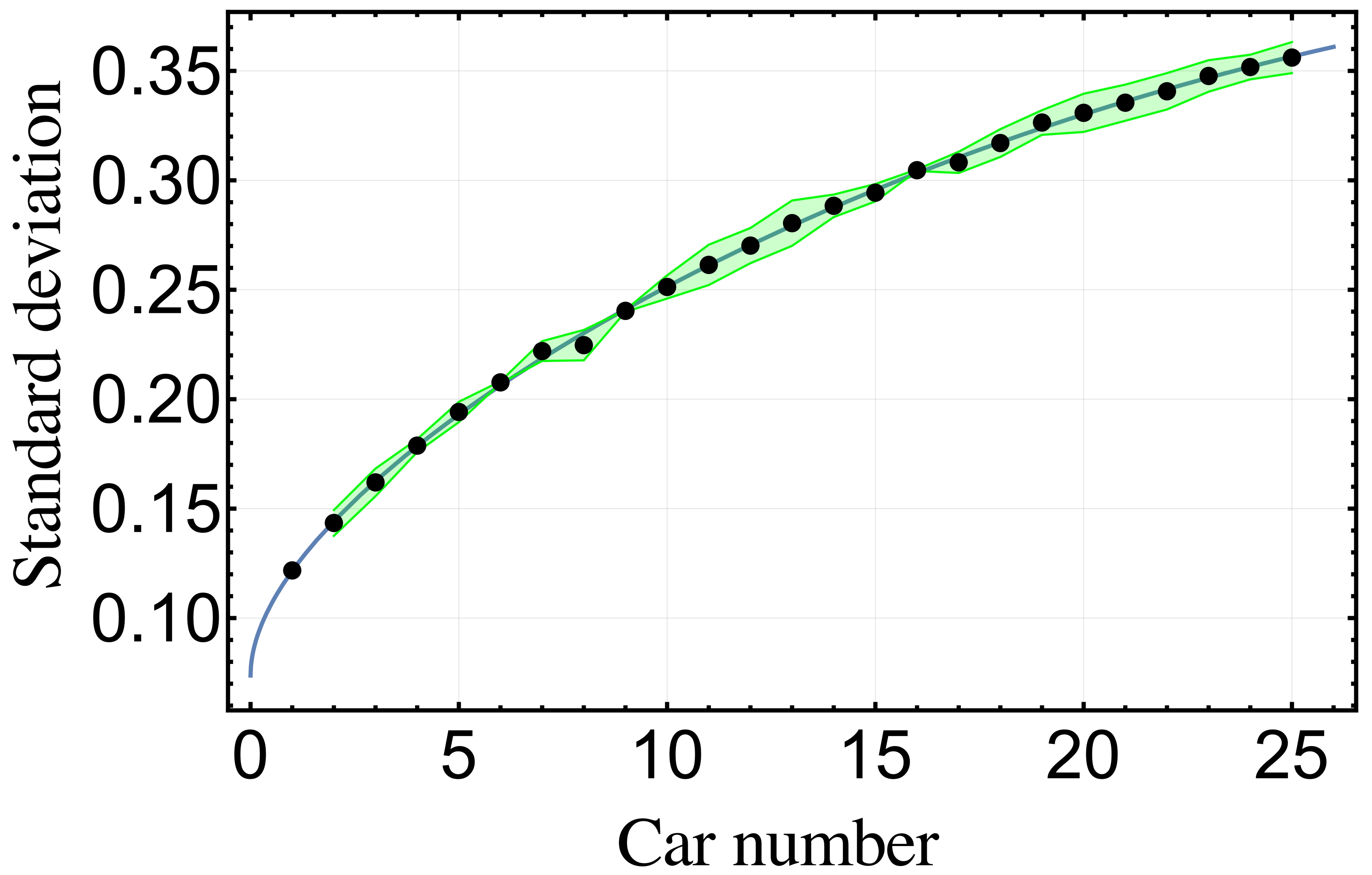}
    \caption{Standard deviation of the time series of the hopping probability of each vehicle for the SOV model.  
    The black points are the simulation results, the green filled band corresponds to the error, and the blue line is the approximate curve.}
    \label{fig:stdv_SOV}
\end{figure}

Furthermore, we confirmed that the SOV model exhibits car-following behavior that is consistent with empirical experiments \cite{jiang2014}. As a vehicle in a platoon is positioned further behind the leading vehicle, there is a concave increase in the standard deviation of the time series of its velocities, revealed for the first time by the empirical experiments in \cite{jiang2014}. This characteristic is not reproduced by traditional deterministic car-following models, such as the optimal velocity model. 
The details of the simulation are as follows:
We started by placing a platoon of 25 vehicles on a 1000-cell circuit. While the leading vehicle accelerated and its hopping probability reached 1.0, we evaluated the standard deviation of the time series of velocity for each vehicle. To ensure accuracy, we conducted this simulation three times and used the average of the three standard deviations for the plot shown in Figure \ref{fig:stdv_SOV}. In Figure \ref{fig:stdv_SOV}, the black points represent the means of the three standard deviations for each vehicle, and the green filled band corresponds to the error of the evaluated standard deviations.

\subsection{Controlled vehicles in the CSOV model}

In this paper, \add{controlled vehicles} (CVs) aim for the ideal velocity at each time step through controls; this ideal velocity is referred to as the \textit{target velocity}. The CV's velocity is calculated as a weighted average between the velocity of the HV as given in Equation \eqref{eq: sov} and the target velocity specified by a control strategy.
Here, we propose two vehicle-control strategies for the CSOV model: the gap-based control (GC) strategy and the flow-based control (FC) strategy. Hereinafter, we refer to vehicles with the GC and the FC as GCVs and FCVs, respectively. 

To ensure that the velocity $v_i^{t+1}$ remains within the range $[0,1]$ of stochastic models, we introduce a weight parameter $\beta$ that regulates the proportion of the two factors: the velocity given in Equation \eqref{eq: sov} and the target velocity.
Thus, we define the velocity of the next time step, given as a weighted average in the following form
\begin{align}
    v_i^{t+1} &=  (1-\beta) \Bigl((1-\alpha)v_i^t + \alpha V(\Delta x_i^t)\Bigr) + \beta C(\cdot),
    \label{eq:csov}
\end{align}
where $C(\cdot)$ represents a function that determines the target velocity 
by each of the controls as described in the following sections.
In the case when $\beta=0$, which corresponds to the scenario where no control is applied, Equation \eqref{eq:csov} is reduced to the SOV model, as represented by Equation \eqref{eq: sov}. Hereinafter, we denote the argument of the control function $C(\cdot)$ as $\Delta\Tilde{x}_i^{t+1}$, which is the gap for the controls, meaning the target gap at the next time step.

\subsection{Gap-based control (GC)}\label{subsec:GC}
The GC strategy is a control strategy for regulating the motion of a GCV by smoothing the gap between the front and rear of the GCV to reduce the unevenness of the distances between the vehicles, which is based on the idea of \cite{zhu2018}. The target velocity given by the control function $C(\cdot)$ is defined as the value of the optimal velocity function evaluated as the mean of the front and rear gaps:
\begin{align}
C_G(\Delta \tilde{x}_{i}^{t+1}) &= V(\Delta \tilde{x}_i^{t+1})=V\left(\frac{\Delta x_{i}^t + \Delta x_{i-1}^t}{2}\right)
\label{eq:GC},
\end{align}
where $\Delta x_{i}^t$ and $\Delta x_{i-1}^t$ denote the front and rear gaps for the $i$th vehicle at time step $t$, respectively, and $\Delta \tilde{x}_i^{t+1}$ is the mean of these two gaps, which can be a rational number. The variable of the OV function can be accepted to have rational numbers as its arguments. Note that, to distinguish the control strategies clearly, the control function $C(\cdot)$ for the GC strategy is denoted as $C_G(\cdot)$.

\subsection{Flow-based control (FC)}\label{subsec:FC}
We define the local flow first to introduce the FC strategy. The local flow of the $i$th vehicle $Q_i^t$ at time step $t$ is defined as the product of the local density $\rho_i^t$ and its velocity $v_i^t$ in terms of the gap $\Delta x_i^{t}$:
\begin{align}
 Q_i^{t}= \rho_i^{t}v_i^{t}
 =\frac{1}{\Delta x_i^{t}+1}v_i^{t}.
\label{eq:localflow_def}
\end{align}
Note that the density used in the calculation of the local flow is the reciprocal of the front gap. 
To determine the velocity at the next time step, we assume that an FCV will adjust its velocity so that its local flow at the next time step will be equal to the average flow of the vehicles in front $Q_{i}^t$ and behind it $Q_{i-1}^t$:
\begin{align}
Q_i^{t+1}= \frac{Q_i^t+Q_{i-1}^t}{2}. 
\label{eq:averageflow}
\end{align}
By the definition of the local flow in Equation \eqref{eq:localflow_def}, the FCV can reach the ideal flow with the gap $\Delta\tilde{x}_i^{t+1}$:
\begin{align}
 Q_i^{t+1}
 =\frac{1}{\Delta\tilde{x}_i^{t+1}+1}V(\Delta\tilde{x}_i^{t+1}),
\label{eq:localflow}
\end{align}
where $\Delta\Tilde{x}_i^{t+1}$ can be obtained by solving Equation \eqref{eq:localflow} numerically, since the left-hand side of Equation \eqref{eq:localflow} can be calculated by the flow rate at time step $t$ from Equation \eqref{eq:averageflow}.
Provided that the gap $\Delta\Tilde{x}_i^{t+1}$ is given, the target velocity for the FCV is determined by evaluating the optimal velocity function, that is,
\begin{align}
\begin{split} 
    C_F\left(\Delta \tilde{x}_i^{t+1}\right)=V\left(\Delta \tilde{x}_i^{t+1}\right).
    \label{eq:control_FC}
\end{split}
\end{align}
Similarly, the control function $C(\cdot)$ for the FC strategy is denoted as $C_F(\cdot)$.

\begin{figure}[ht]
    \centering
     \includegraphics[width=0.7\linewidth]{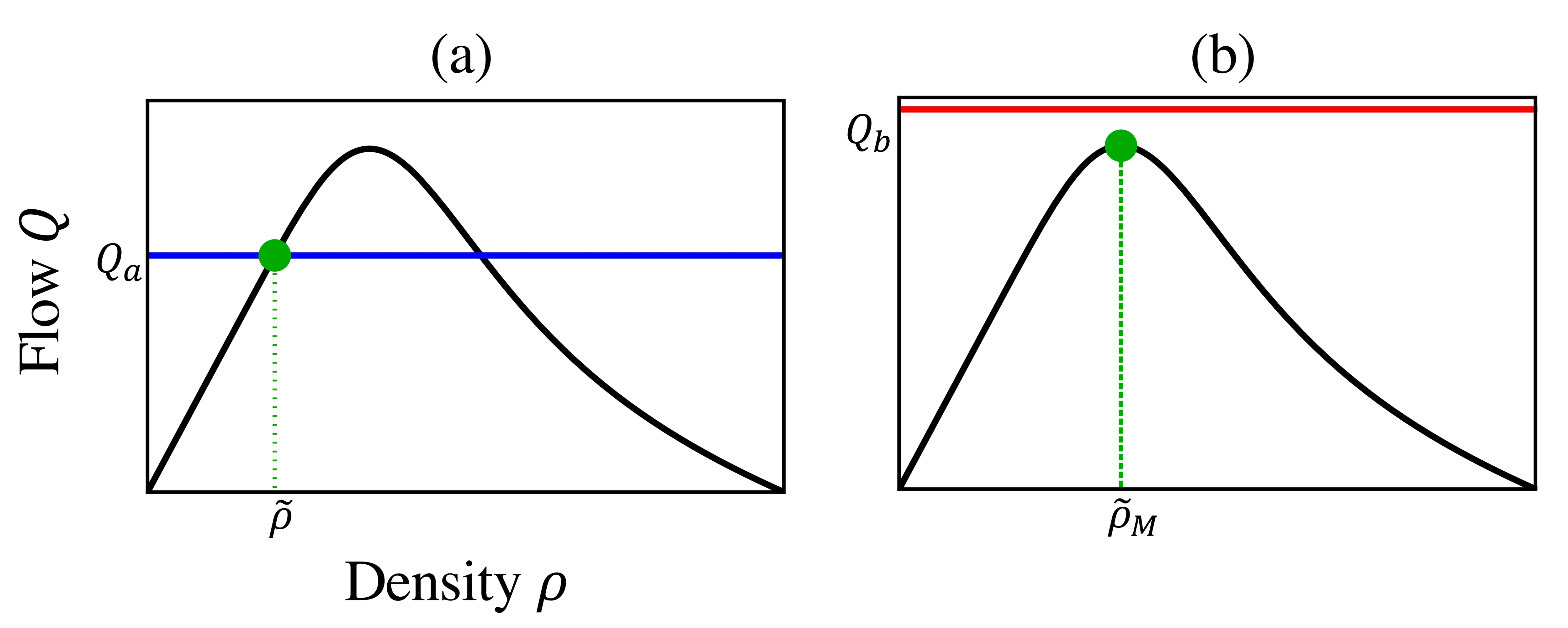}
     \caption{Two cases of Equation \eqref{eq:localflow}. (a) When there are two points of intersection, the left-side intersection is adopted for determining $\Delta \tilde{x}_i^{t+1}=-1+1/\tilde{\rho}$. (b) When there is no point of intersection, we adopt $\Delta \tilde{x}_i^{t+1}$ as $-1+1/\tilde{\rho}_M$.}
    \label{fig:right_intersection}
\end{figure}

We will remark on the choice of the gap $\Delta\tilde{x}_i^{t+1} = -1+1/\tilde{\rho}$ determined by solving Equation \eqref{eq:localflow} of the density $\tilde{\rho}$. 
\add{There are} \erase{The equation leads to} two possible cases, depending on the flow \erase{$Q$} \add{$Q_i^{t+1}$} on the left-hand side of the equation. The first case is when there are any solutions to the equation, and the second case is when there is no solution. 
Figure \ref{fig:right_intersection} (a) and (b) schematically depict the density to be chosen with the green points for the above two cases. 
In the former case, a line $Q=Q_a$ on the left-hand side of Equation \eqref{eq:localflow} meets the curve in terms of $\rho$ on the right-hand side of Equation \eqref{eq:localflow}, as shown in Figure \ref{fig:right_intersection} (a). 
Though we have two points of intersection, we choose the left-side intersection $\rho = \tilde{\rho}$ to calculate the gap. 
This is because it is more reasonable to aim for a low density and free flow condition than a high density and jamming flow condition to achieve the same flow rate.
In the latter case, we choose the density $\tilde{\rho}_M$ that maximizes the right-hand side to calculate the gap.

\section{Simulations}
We conducted computer simulations of complete CV and mixed traffic flow. The initial positions of the vehicles were randomly assigned on a one-lane and 200-cell circuit with the periodic boundary condition; once the number of vehicles on the circuit was determined, no vehicles were allowed to enter or exit the circuit, that is, the global density of the system is fixed. The simulations were terminated after 50,000 time steps. 
As depicted in Figure \ref{fig:OVfunction}, the rear gap for the $i$th vehicle is $\Delta x_{i-1}^t$. As we are using periodic boundary conditions, the front gap of the first vehicle and the rear gap of the last vehicle are the same. Denoting the number of vehicles by $N$, this gap is given by $x_1^t-x_N^t+L-1$.

To investigate the impact of vehicle placement on the performance of the control strategies, we considered two different patterns of CV placement for the mixed traffic flow: uniform placement, in which the CVs were evenly distributed among vehicles on the circuit, and block placement, in which the CVs were grouped together as a \textit{block}.

\subsection{Complete CV traffic}
\begin{figure}[h]
 \begin{center}
    \includegraphics[width=0.9\linewidth]{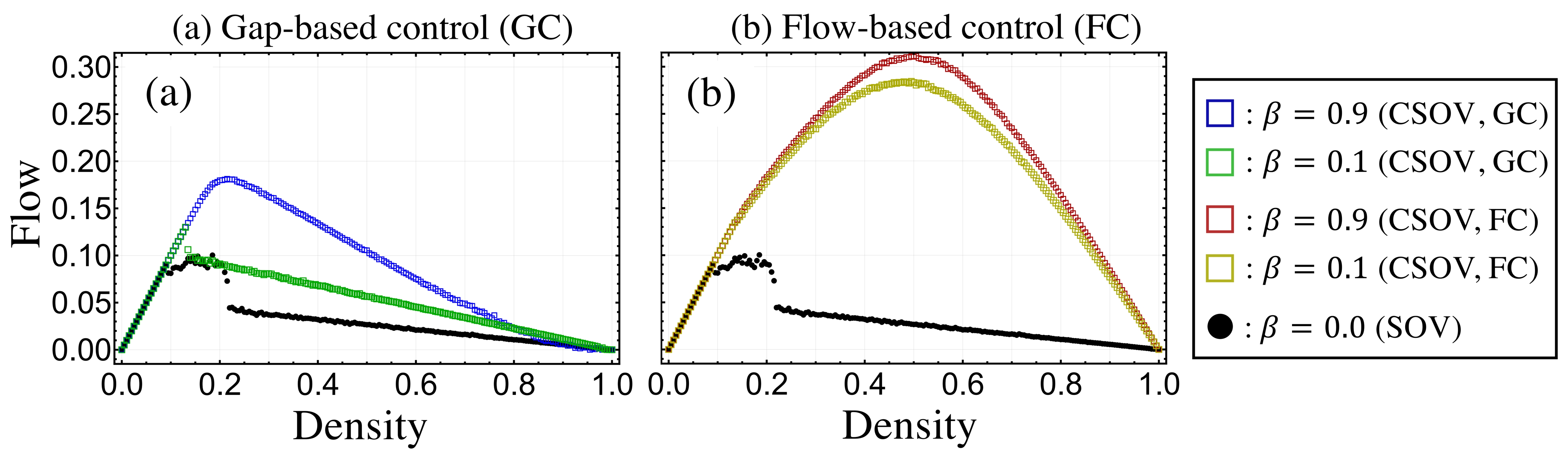}
  \caption{Fundamental diagrams for the complete CV traffic. (a) Fundamental diagram when the CVs are implemented on the GC strategy. (b) Fundamental diagram when the CVs are implemented on the FC strategy. The black points represent the fundamental diagram for the SOV model for comparison.}
  \label{fig:FD_completeCV}
 \end{center}
\end{figure}

\begin{figure}[h]
 \begin{center}
  \includegraphics[width=0.85\linewidth]{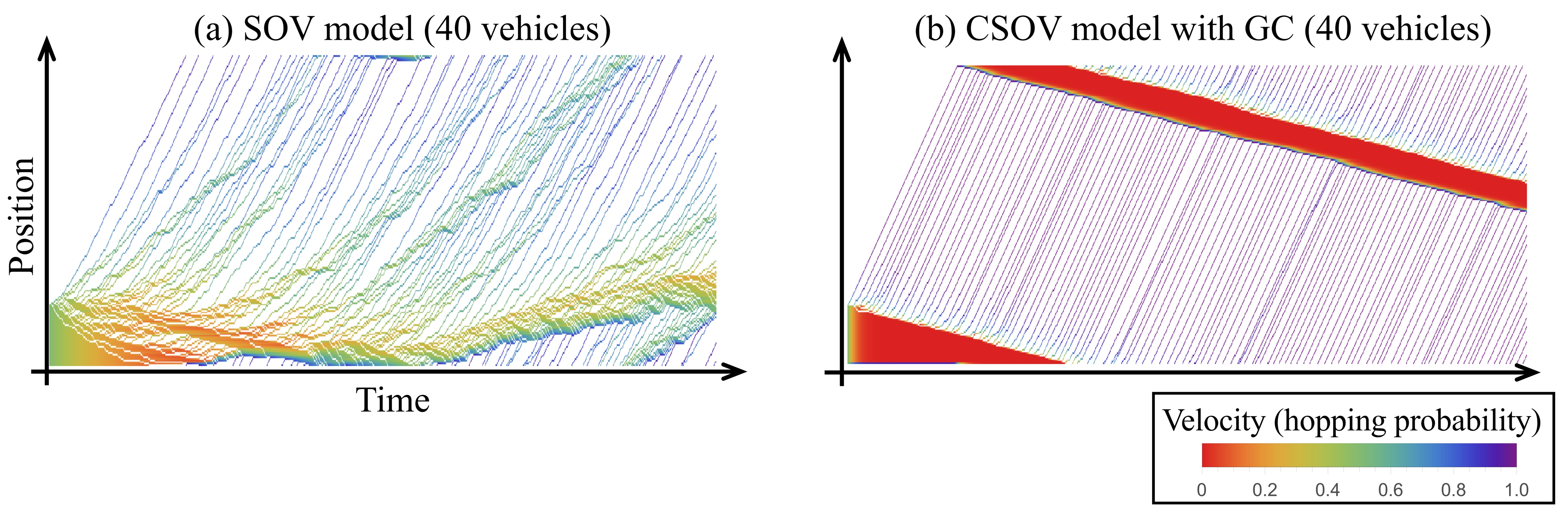}
  \caption{Time--space diagrams for (a) SOV model and (b) CSOV model with the GC strategy when the density $\rho=0.2$. The vehicles are placed as a platoon.}
  \label{fig:addtional_exp}
 \end{center}
\end{figure}

Figures \ref{fig:FD_completeCV} (a) and (b) show the fundamental diagrams of complete CV traffic cases where the GC and FC strategies are applied, respectively. In each panel, two scenarios when the control parameter $\beta=0.1$ (weak control effect) and 0.9 (strong control effect) are plotted with empty squares, along with the fundamental diagram of the SOV model with black points for comparison. In the free flow state when the density $\rho \in [0,0.1]$, the flow of both control strategies coincides with that of the SOV model regardless of the strength of the control parameter $\beta$ because each term in Equation \eqref{eq:csov} approaches 1 in the free flow state. 
In the other densities, the complete CV traffic with both control strategies results in higher flow than the SOV model, except in the case of the GC strategy with a weak control parameter $\beta=0.1$, where the density \erase{$\rho\in[0.17,0.25]$}\add{$\rho\in[0.135,0.2]$}.

In that density region \erase{$\rho\in[0.17,0.25]$}\add{$\rho\in[0.135,0.2]$}, \erase{the SOV model surprisingly shows higher flow than}\add{the flow in the SOV model belongs to metastable state, although the flow in} the CSOV model with the GC strategy for $\beta=0.1$ \add{is already in the jamming state, as} plotted in the green squares in Figure \ref{fig:FD_completeCV} (a). 
This is likely due to the GCVs slowing down when the GCVs leave the region of traffic congestion. 
We conducted an additional simulation to observe such behavior of the GCVs. Figures \ref{fig:addtional_exp} (a) and (b) show the time--space diagrams for the SOV and CSOV models with the GC strategy, respectively, when the density $\rho=0.2$. The time--space diagrams depict the trajectory of each vehicle, with the color indicating the hopping probability. In this additional simulation, the vehicles are placed in a platoon with velocity $v=0.5$ at the initial time. First, the velocities of the GCVs in the platoon decrease to around 0, indicated in red in Figure \ref{fig:addtional_exp} (b) because the GCVs take account of both front and rear gaps. After the GCVs leave the traffic congestion, they drive faster, maintaining sufficient gaps compared to the HVs. However, as shown in Figure \ref{fig:addtional_exp} (b), the traffic congestion remains even at the time when the congestion is eased in the SOV model. 

A similar interpretation can be applied to the fundamental diagram shown in Figure \ref{fig:FD_completeCV} (a) based on the additional simulation. 
When a GCV is located in front of a traffic congestion, since the vehicle behind, which is stuck in the traffic congestion, drives close to the GCV, the rear gap of the GCV is considerably smaller than its front gap between the GCV and the vehicle in the front, which is not stuck in the traffic congestion. Notably, a GCV determines its velocity considering both front and rear gaps, whereas an HV determines its velocity only by its front gap. Thus, the GCV is affected by its rear gap and decelerates.

In contrast, the CSOV model with the FC strategy improved the flow in the metastable and congestion state, as shown in Figure \ref{fig:FD_completeCV} (b), even when the effect of the control is weak such as $\beta=0.1$. Moreover, the FC strategy achieved a higher flow than the GC strategy, as shown by the two panels in Figure \ref{fig:FD_completeCV}. The shape of the fundamental diagram for the complete CV traffic, shown in Figure \ref{fig:FD_completeCV} (b), is different from that of the measured fundamental diagram in Figure \ref{fig:FD_comp} (b) because the FCVs are not affected by the velocities of their rear vehicles and thus, exhibit almost constant velocities.

\subsection{Mixed traffic}
To investigate the impact of CV on mixed traffic flows, we conducted the simulations for the CV penetration rates from 10\add{\%} to 90\% by 10\%. The results indicate that even with a low penetration rate of 30\% and weak control ($\beta=0.1$), the traffic flow is improved compared to that of the SOV model. Moreover, the characteristics of each control strategy are evident in the fundamental diagrams and remain consistent up to 80\% penetration rate. Beyond 80\% penetration rate, the mixed traffic exhibits similar behavior to that of complete CV traffic. Figure \ref{fig:FD_30,50percentCV} (a), (b),\erase{ and} (c)\add{, and (d)} display the fundamental diagrams for CV penetration rates of 30\%, \add{40\%,} 50\%, and 70\%, respectively.
The shape of the data points indicates the placement of the CVs: squares represent block placement, and circles represent uniform placement. 

\begin{figure}[h]
 \begin{center}
  \includegraphics[width=0.9\linewidth]{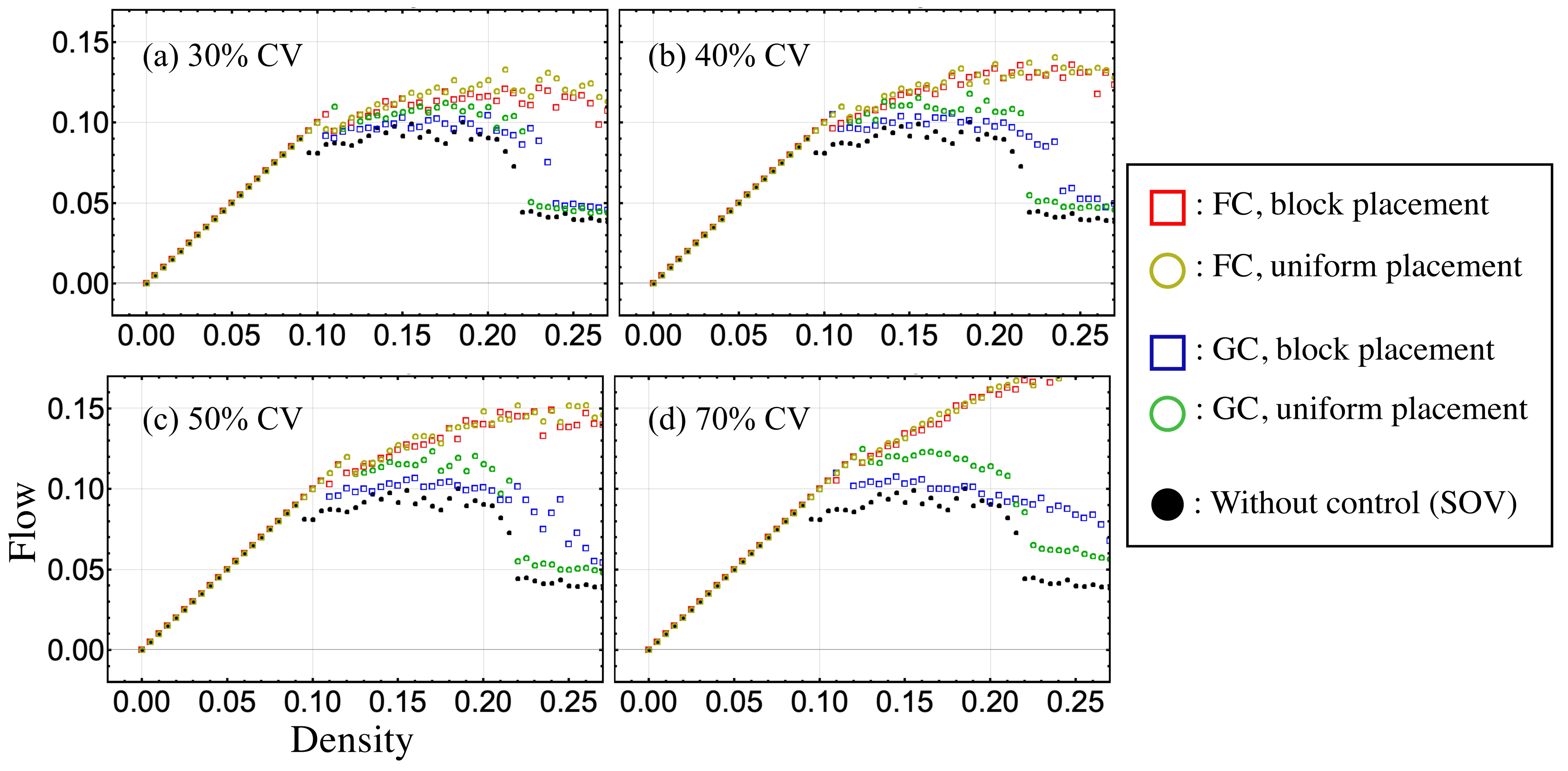}
  \caption{Fundamental diagram of the mixed traffic flow with CVs and HVs at different penetration rates of CVs. For both controls, the control parameter $\beta$ is set as 0.1. The black points represent the SOV model case for comparison.}
  \label{fig:FD_30,50percentCV}
 \end{center}
\end{figure}

Our simulations revealed that the placement of GCVs affects flow.
For all CV penetration rates in Figure \ref{fig:FD_30,50percentCV}, the block placement of the GC strategy extends the density range of the metastable state to higher densities by comparing the uniform placement of the GC strategy as observed in Figure \ref{fig:FD_30,50percentCV} (a), (b),\erase{ and} (c)\add{, and (d)}. While the uniform placement of the GC strategy does not extend the range of the metastable state, it results in higher flow in the metastable state compared to the SOV model and the block placement of the GC strategy. 

Conversely, the flow of the mixed FCV traffic remains independent of their placement, yielding almost identical flow rates. This is illustrated by the red squares and yellow circles in Figure \ref{fig:FD_30,50percentCV} (a), (b),\erase{ and} (c)\add{, and (d)}. The presence of FCVs in the mixed traffic slightly extends the range of densities for the free flow state more than that of GCVs. Particularly, the end density of the free flow state is approximately \erase{$\rho\sim 0.115$ and 0.125}\add{between 0.1 and 0.12} for penetration rates of 30\%\add{, 40\%} and 50\%, respectively, compared to \erase{$\rho=0.1$}\add{$\rho=0.09$} in the SOV model.
At a 70\% penetration rate, discerning the boundary of the free flow state is challenging. The flow, in this case, increases linearly within the free flow state density range and smoothly connects to the flow of the metastable state. The FC strategy expands the range of densities for the metastable state. Furthermore, it provides a higher flow rate compared to the SOV model at 30\%\add{, 40\%} and 50\% penetration rates.

\begin{figure}[h]
    \centering
    \includegraphics[width=0.85\linewidth]{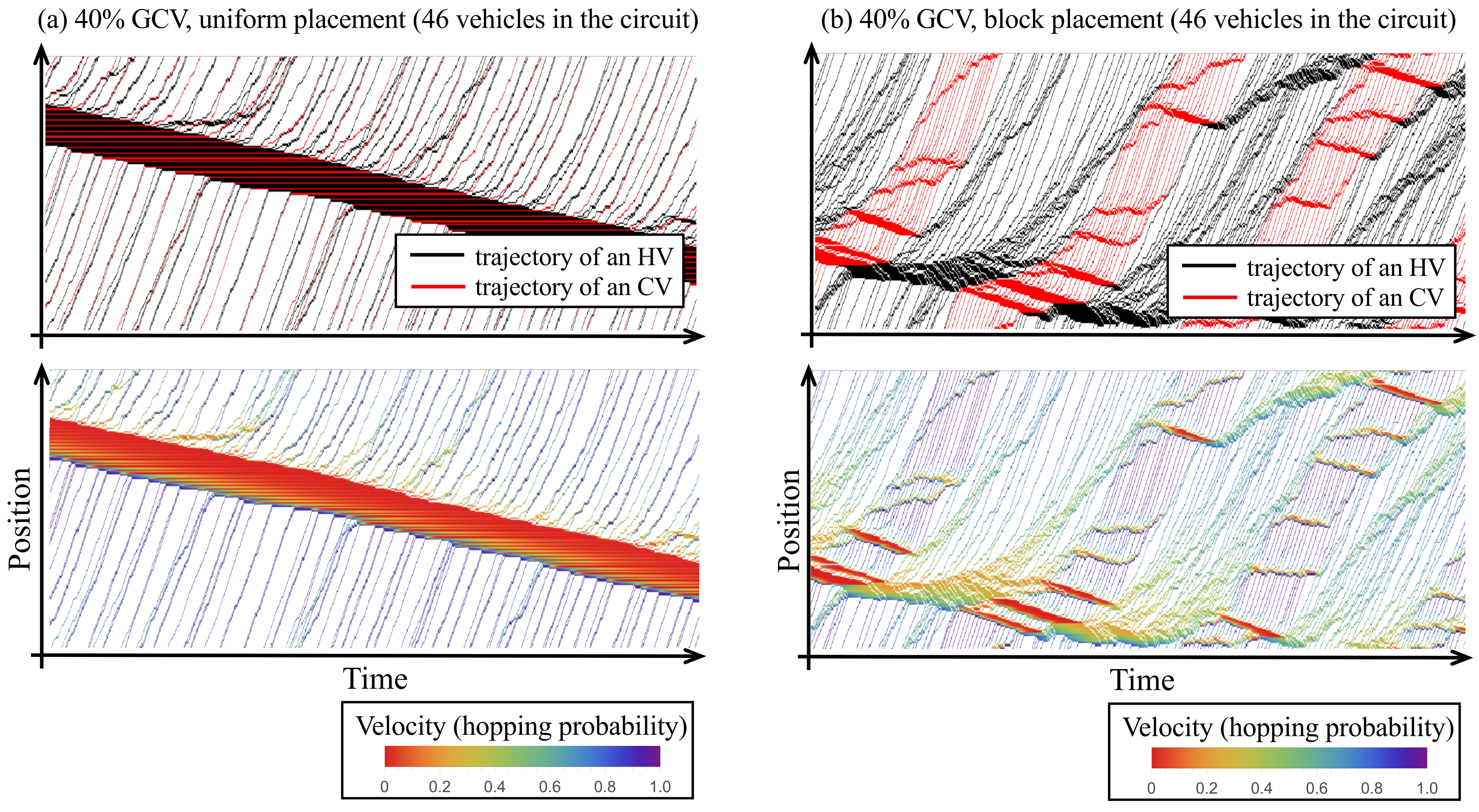}
    \caption{Time--space diagrams for mixed traffic flow with GCVs and HVs when GCVs are put (a) uniformly and (b) in a block. The number of vehicles is \erase{50}\add{46} on the circuit, and the penetration rate of the GCVs is \erase{30}\add{40}\%. }
    \label{fig:TS_GC}
\end{figure}

\begin{figure}[h]
    \centering
    \includegraphics[width=0.85\linewidth]{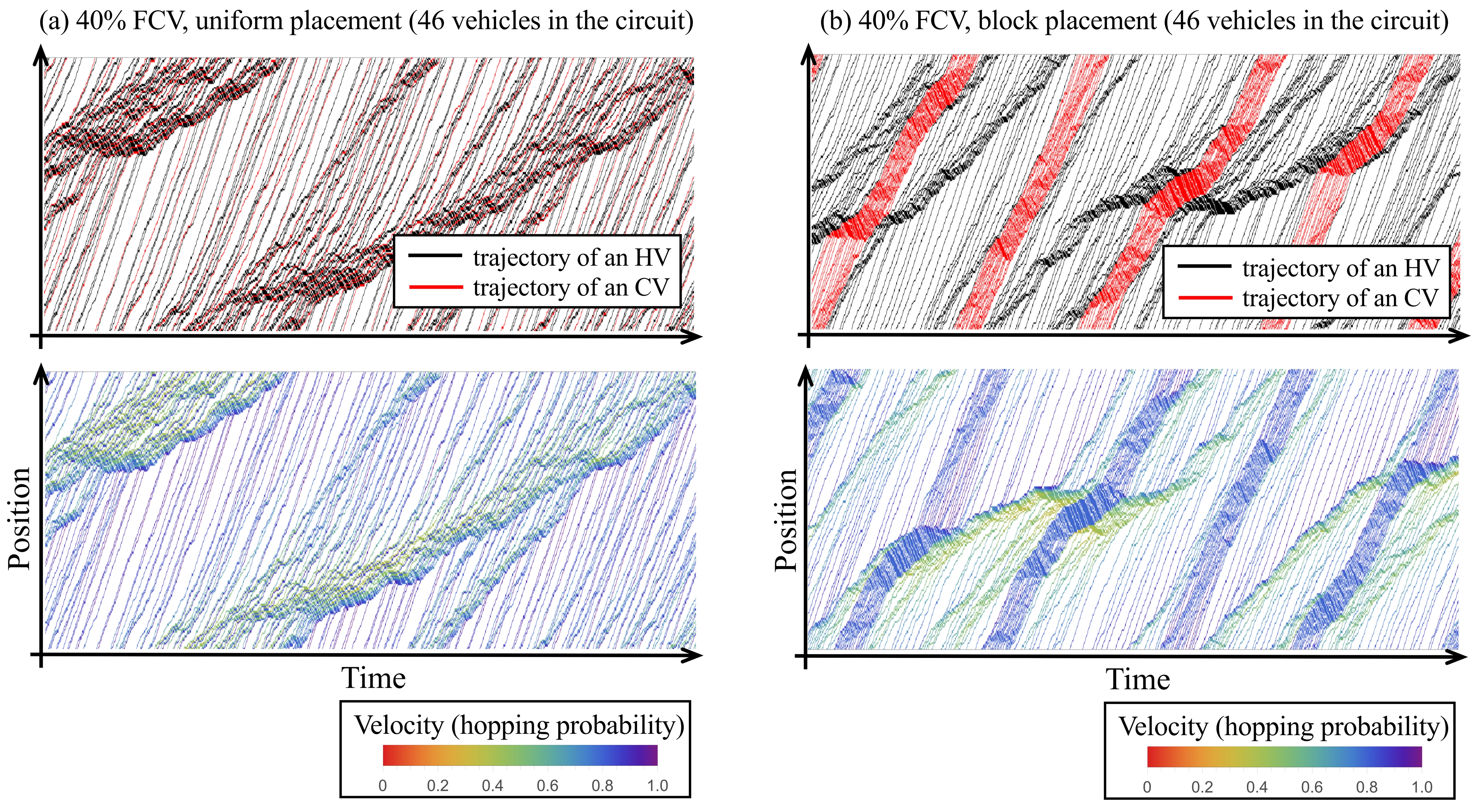}
    \caption{Time--space diagrams for mixed traffic flow with FCVs and HVs when FCVs are put (a) uniformly and (b) in a block. The number of vehicles is \erase{50}\add{46} on the circuit, and the penetration rate of the FCVs is \erase{30}\add{40}\%.}
    \label{fig:TS_FC}
\end{figure}

To further analyze these results, we plotted time--space diagrams to visualize the flow obtained from our simulations. Figure \ref{fig:TS_GC} and \ref{fig:TS_FC} present the time--space diagrams for a scenario in which \erase{50}\add{46} vehicles are on the circuit, and \erase{30}\add{40}\% of them are controlled, which corresponds to the density of \erase{$\rho = 0.25$}\add{$\rho = 0.23$} in Figure \ref{fig:FD_30,50percentCV} \erase{(a)}\add{(b)}. At this density, the flow of the block GCV placement is in the metastable state, whereas the flow of the uniform GCV placement is in the jamming state. The time--space diagrams depict the trajectory of each vehicle, with the color indicating the hopping probability. While a single large congestion is present in the uniform placement (see Figure \ref{fig:TS_GC} (a)), the scenario of block placement with GC in Figure \ref{fig:TS_GC} (b) results in higher flow compared to the uniform placement. This is due to the formation of smaller congestions at various locations, which prevent the formation of a single large congestion. These small congestions are caused solely by GCVs. Conversely, in the case of the FC strategy, FCVs do not participate in the congestion, resulting in the absence of congestion, as seen in Figure \ref{fig:TS_FC}.

\subsubsection{Comparison of the two controls under similar performance} 
\begin{figure}[h]
    \centering
    \includegraphics[width=0.85\linewidth]{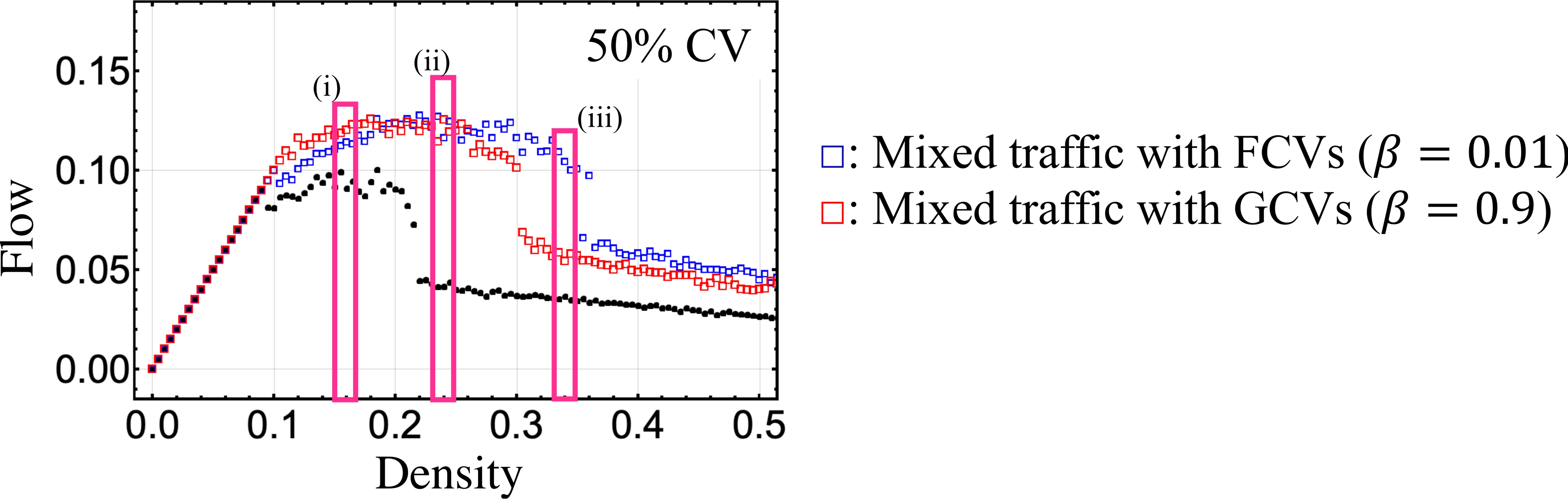}
    \caption{Fundamental diagram for the HVs traffic (black points) and the mixed traffic with CVs and HVs (squares). The red empty squares represent the case when the CVs are GCVs, and the blue ones represent the case when the CVs are FCVs.}
    \label{fig:FD_001FCvs09GC}
\end{figure}

Earlier, we observed that the flow of traffic with FCVs was higher than that with GCVs for the same control parameter $\beta$. However, it remains unclear whether there are any differences between the two control strategies when they perform \textit{similarly};
that is, the fundamental diagrams for the FC and the GC strategies show almost the same peak flow at the same density.
The FC for $\beta=0.01$ plotted in blue squares and the GC for $\beta=0.9$ plotted in red squares have similar shapes of fundamental diagrams for a mixed traffic system comprising 50\% HVs and 50\% CVs placed in a block formation (Figure \ref{fig:FD_001FCvs09GC}). Both control strategies produced higher flow rates than the SOV model for \erase{$\rho > 0.1$}\add{$\rho \geq 0.1$}.
When comparing the FC and the GC cases, we identified three distinct density regions: (i) a region where the GC performed slightly better than the FC, (ii) a region where both controls performed similarly, and (iii) a region where the FC was in a metastable state but the GC was in a jamming state. 

We conducted a detailed analysis to investigate the traffic further by plotting time--space diagrams and velocity distributions for each region. When plotting the velocity distributions, we specifically considered velocities at a fixed position during the time interval $t \in [5000, 6500]$, representing a period when the traffic had reached a steady state.

\begin{figure}[h]
    \centering
    \includegraphics[width=0.85\linewidth]{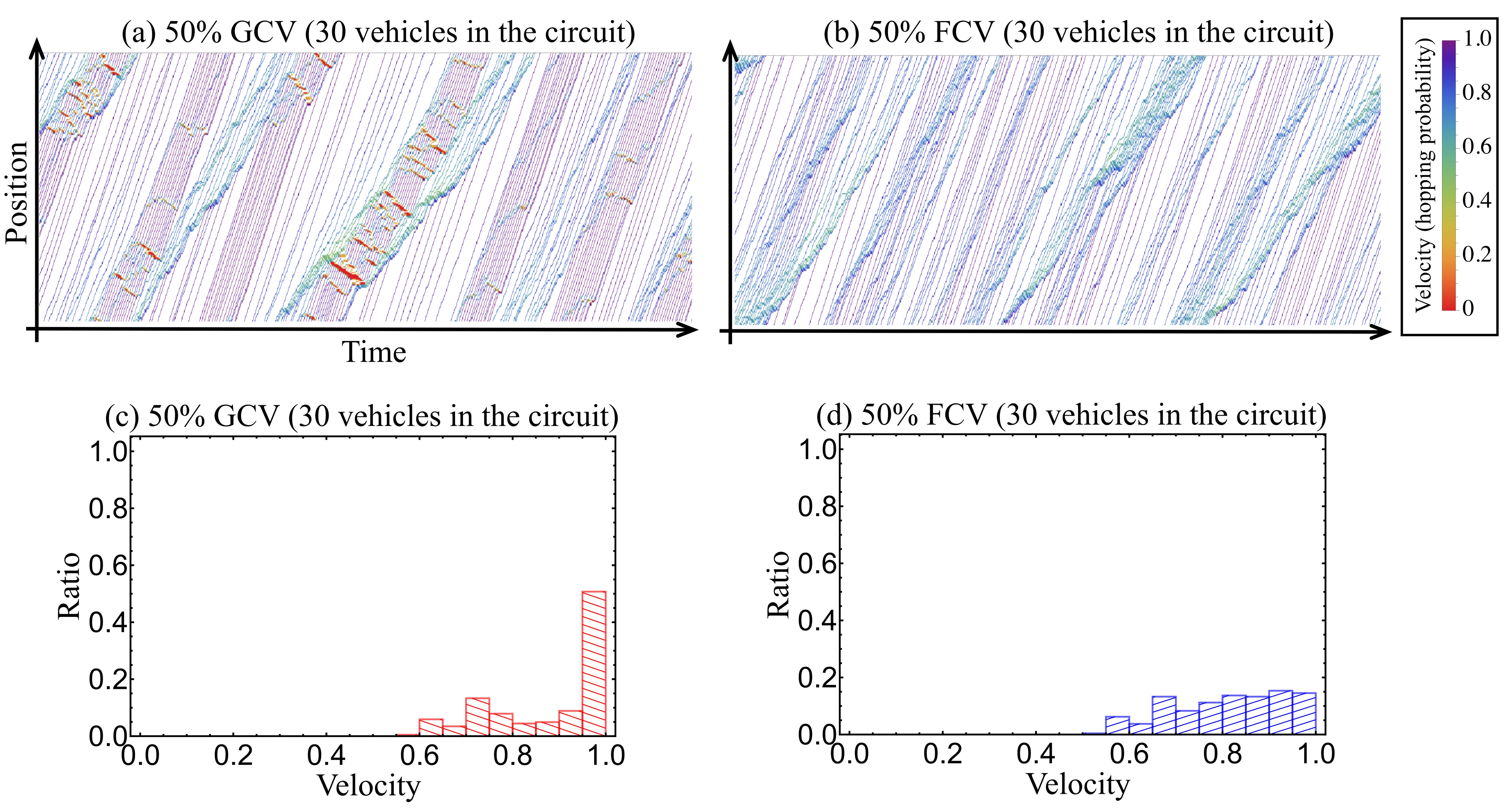}
    \caption{Comparison of the mixed GCV traffic and mixed FCV traffic. The CVs with both control strategies are placed in a block, and the total number of the vehicles is 30, which is in the region (i) of Figure \ref{fig:FD_001FCvs09GC}.
    Time--space diagrams when the CVs use (a) GC strategy and (b) FC strategy. Velocity distribution of the vehicles in the circuit for (c) GCV mixed traffic and (d) FCV mixed traffic.}
    \label{fig:ts_histo_N30}
\end{figure}

In region (i), as indicated by Figure \ref{fig:FD_001FCvs09GC}, the GC leads to a higher flow compared to the FC. The corresponding time--space diagram in Figure \ref{fig:ts_histo_N30} (a) reveals that the GCVs drive with their velocity $v\sim 1$, indicated by purple lines with small gaps in the time--space diagram. The driving strategy of the GCVs gives the HVs more space than that of the FCVs, as shown in Figure \ref{fig:ts_histo_N30} (b), resulting in a higher total flow on the circuit. The velocity distribution in Figure \ref{fig:ts_histo_N30} (c) reflects this result; for the GCV mixed traffic, 60\% of the vehicles on the circuit passing through a cell are in the velocity range $v\in[0.95,1.0]$. Conversely, for the mixed FCV traffic case, although the mode of the velocity distribution is $v\in[0.95,1.0]$, the velocity is uniformly distributed in the range \erase{0.7}\add{0.65} to 0.95. This wider velocity distribution for the mixed FCV traffic results in a lower flow rate than for the mixed GCV traffic.

\begin{figure}[h]
    \centering
    \includegraphics[width=0.85\linewidth]{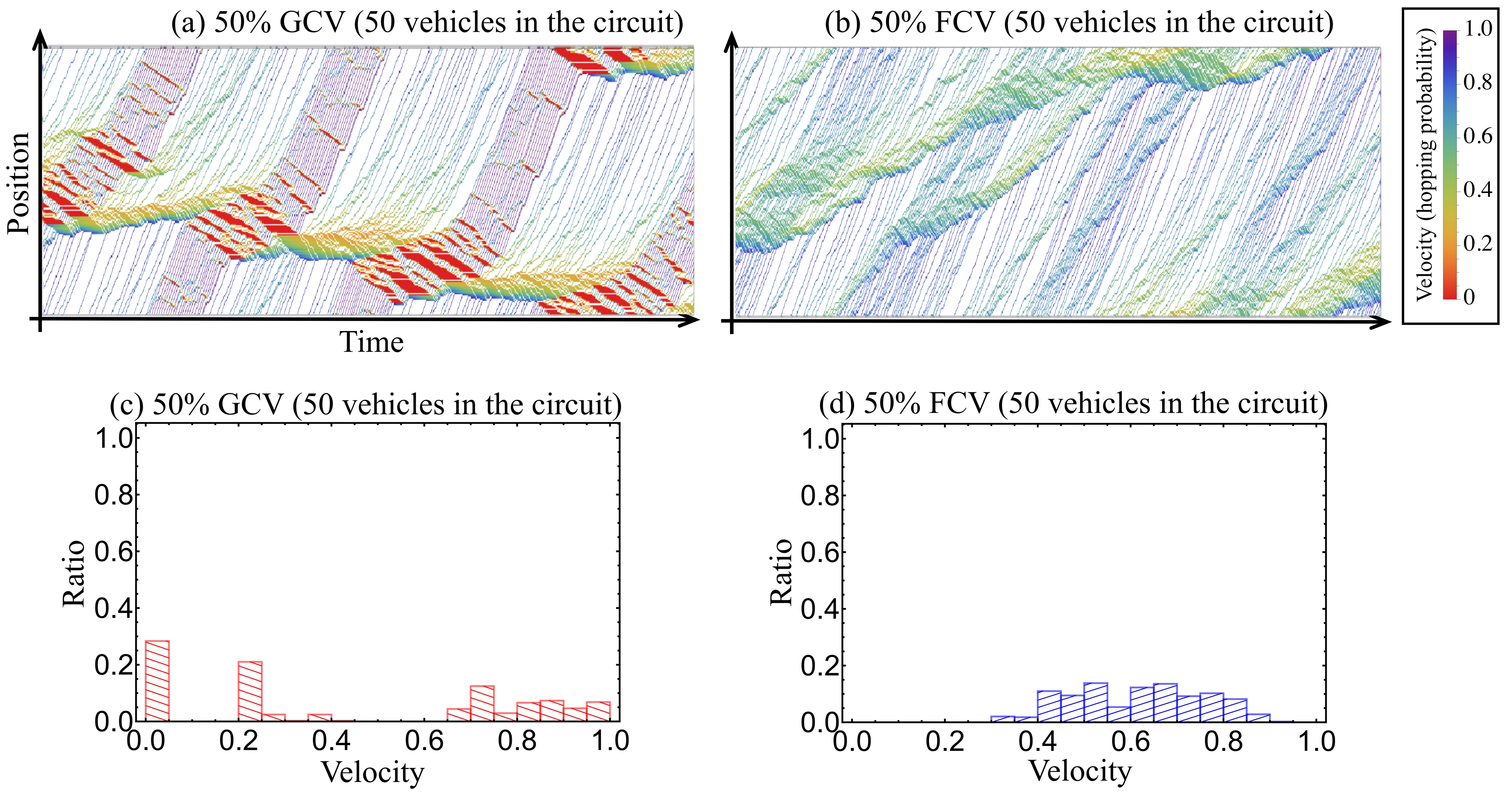}
    \caption{Comparison of the mixed GCV traffic and mixed FCV traffic. The CVs with both control strategies are placed in a block, and the total number of the vehicles is 50, which is in the region (ii) of Figure \ref{fig:FD_001FCvs09GC}.
    Time--space diagrams when the CVs use (a) GC strategy and (b) FC strategy. Velocity distribution of the vehicles in the circuit for (c) GCV mixed traffic and (d) FCV mixed traffic.}
    \label{fig:ts_histo_N50}
\end{figure}

\begin{figure}[h]
    \centering
    \includegraphics[width=0.85\linewidth]{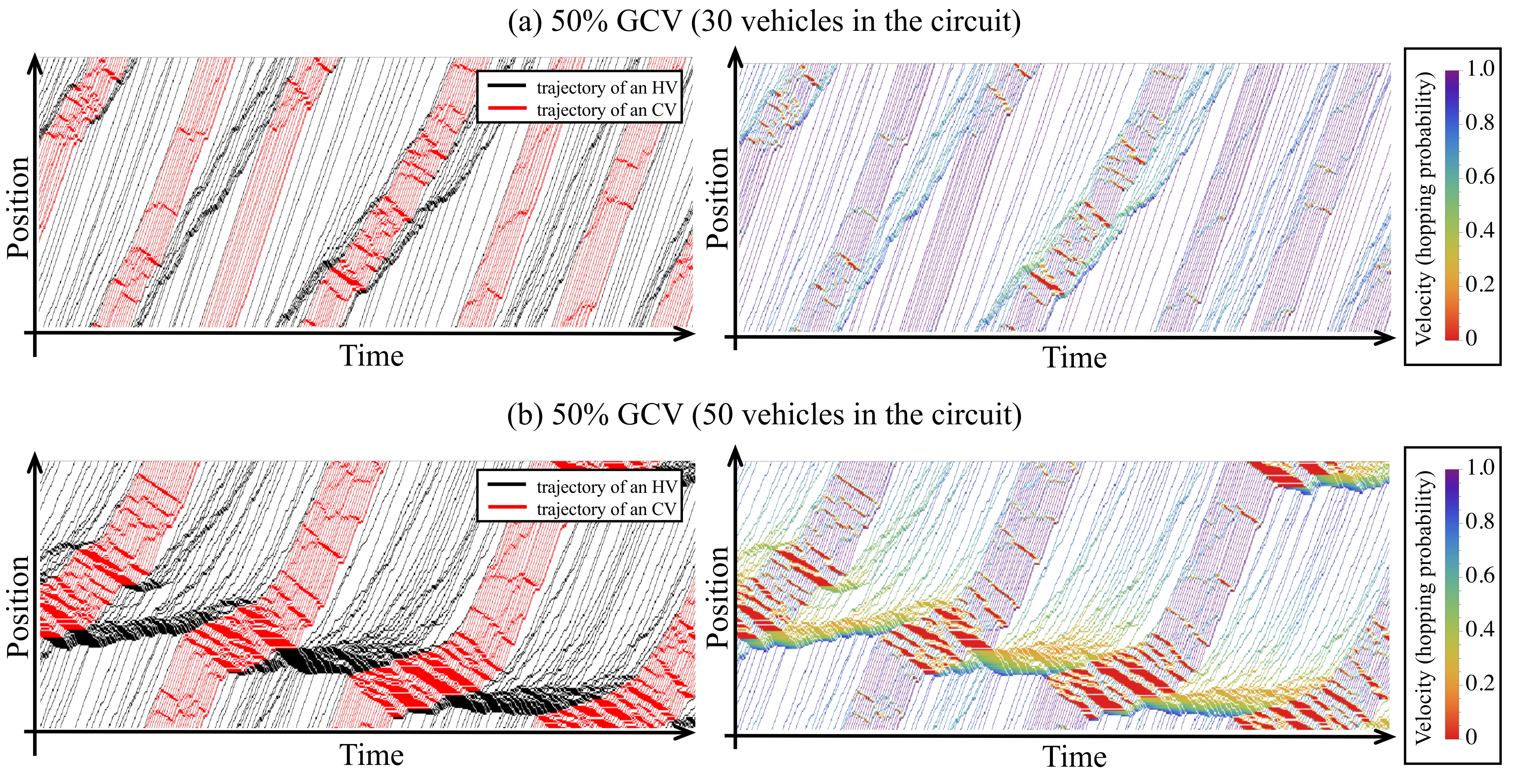}
    \caption{Time--space diagrams for (a) 30 vehicles and (b) 50 vehicles in the circuit. The upper panels identify the HVs (black lines) and the GCVs (red lines). The lower panels of (a) and (b) are identical to Figure \ref{fig:ts_histo_N30} (a) and \ref{fig:ts_histo_N50} (a), respectively. The time--space diagrams of each upper and lower panel are identically the same.}
    \label{fig:specifying_ts}
\end{figure}

However, notably, for the mixed GCV traffic, we observed stop-and-go waves that do not affect all vehicles in the circuit. The traffic congestion is indicated in red narrow bands in Figure \ref{fig:ts_histo_N30} (a) and Figure \ref{fig:ts_histo_N50} (a). 
We replotted those time--space diagrams in the upper panels of Figure \ref{fig:specifying_ts}, which identify the GCVs (red lines) and the HVs (black lines). The lower panels are corresponding time--space diagrams indicating velocity in color. Note that the time--space diagrams of each upper and lower panel are identically the same. As can be seen in the upper panels in Figure \ref{fig:specifying_ts} (a) and (b), the characteristic congestions occur in the block of the GCVs.

These congestions are particularly prominent when the congestion formed by the HVs, indicated in blue or green, passes through the platoon of the GCVs. This observation suggests that the GCVs with a large control parameter, such as $\beta=0.95$, tend to maintain high speeds with minimal front gaps. Consequently, the congestion propagates through stop-and-go waves within the platoon of the GCVs.

Interestingly, even though the flow in the fundamental diagram is almost the same, the distribution of the velocity is apparently different between the two control strategies, as shown in Figure \ref{fig:ts_histo_N50} (c). A wider velocity distribution is associated with less throughput and more fuel consumption in \cite{stern2018}. However, in the velocity distribution for the mixed GCV traffic, the velocities of the vehicles are distributed more widely than in the mixed FCV traffic. For example, the velocity distribution of Figure \ref{fig:ts_histo_N50} (c) shows \erase{10\%}\add{more than 20\%} of vehicles in the circuit have velocity $v\in[0.0, 0.05]$ as the GCVs trapped in the characteristic traffic congestion halt.

\begin{figure}[h]
    \centering
    \includegraphics[width=0.85\linewidth]{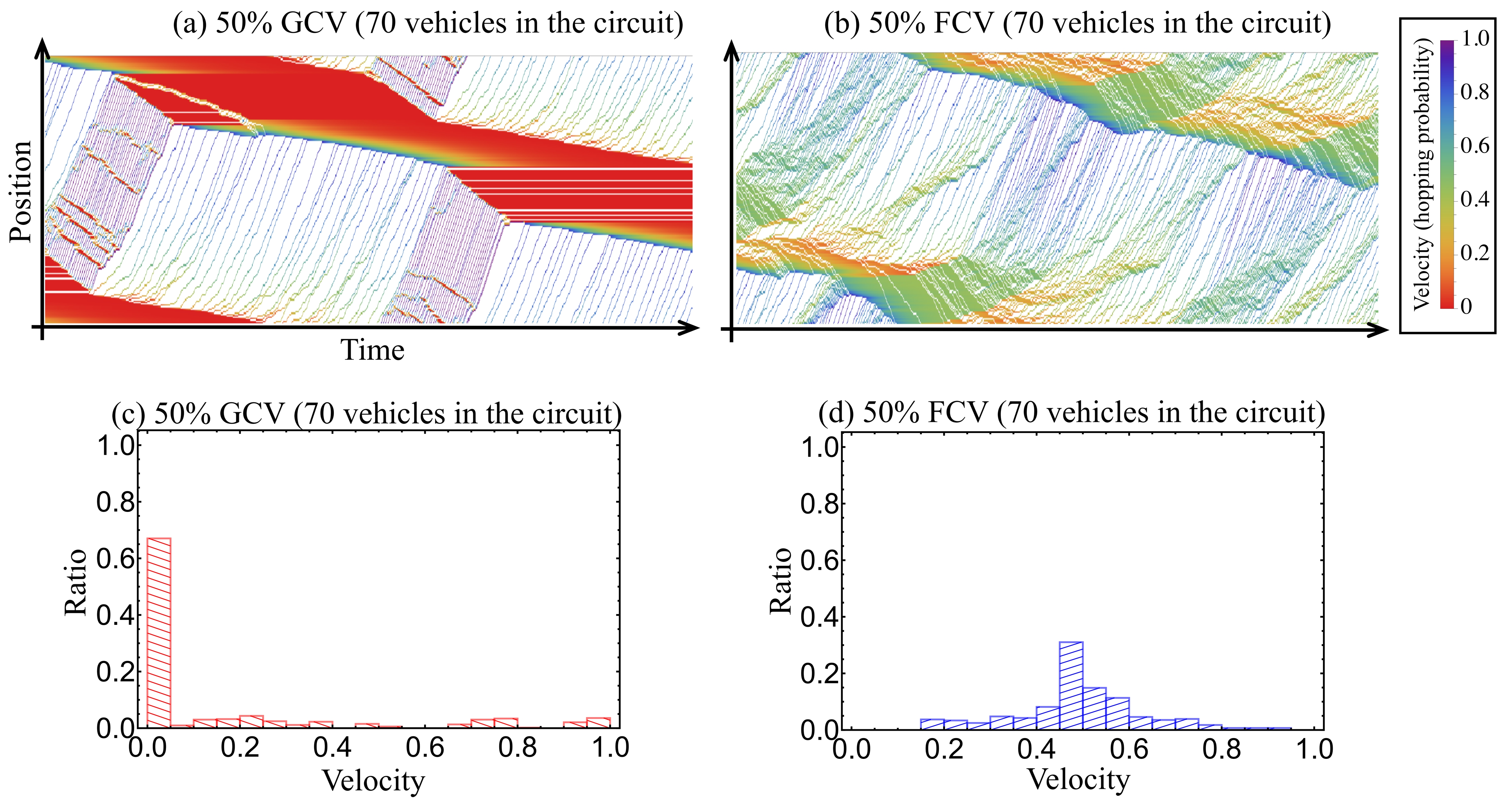}
    \caption{Comparison of the mixed GCV traffic and mixed FCV traffic. The CVs with both control strategies are placed in a block, and the total number of the vehicles is \erase{80}\add{70}, which is in region (iii) of Figure \ref{fig:FD_001FCvs09GC}.
    Time--space diagrams when the CVs use (a) GC strategy and (b) FC strategy. Velocity distribution of the vehicles in the circuit for (c) GCV mixed traffic and (d) FCV mixed traffic.}
    \label{fig:ts_histo_N80}
\end{figure}

In region (iii), the GCVs cannot drive as fast as in regions (i) and (ii); the mixed GCV traffic is in the jamming state as shown in Figure \ref{fig:ts_histo_N80} (a).
Conversely, for the FC case, the median velocity, which is the middle value in the velocity distributions, were recorded as \erase{0.88, 0.74, and 0.56}\add{0.82, 0.62, and 0.47} for regions (i), (ii), and (iii), respectively. These values provide insight into the overall traffic behavior and reveal decreasing median velocities from regions (i) to (iii), suggesting a higher congestion level in region (iii).

\section{Conclusion}

We proposed the CSOV model, which incorporates the effect of vehicle control into the well-known SOV model in the form of a stochastic cellular automaton. By exploiting a notable characteristic of the SOV model, which reproduces the metastable state, we conducted a comprehensive investigation into the effects of vehicle control on both complete CV and mixed traffic scenarios, particularly focusing on the region near the metastable state where the density ranges from $\rho=0$ to \erase{$\rho=0.3$}\add{$\rho=0.25$}. This model enables us to explore more realistic aspects of traffic flow with CVs, including changes in the metastable state and flow characteristics in the presence of CVs.

In this study, we implemented two control strategies within the CSOV model: the GC and FC strategies. The GC strategy regulates a CV's velocity to optimize the smoothing of its front and rear gaps, whereas the FC strategy adjusts a CV's velocity to ensure the smoothness of both the front and rear flow rates. 
When constructing the FC strategy, our focus was on maximizing the flow rate to ease traffic congestion by increasing the number of vehicles passing through per unit of time.

For complete CV traffic, we observed that some GCVs tend to decelerate when leaving a traffic congestion, particularly for weak strength parameter $\beta$. This behavior can be attributed to the influence of the velocities of rear vehicles in the congestion. However, the FC strategy, which considers not only gaps but also velocities, prevents such deceleration when leaving congestion. This difference is evident in the fundamental diagrams when the two control strategies were applied to complete CV traffic, highlighting variations in the density that achieves the maximum flow between the two strategies.

In the case of mixed traffic involving CVs and HVs, we found that regardless of the placement of CVs, the FC strategy expands the region of the metastable state and achieves higher flow in this state compared to the GC strategy. When GCVs are placed in a block, the formation of localized but prominent congestions solely caused by GCVs prevents the emergence of a single large congestion in the system. Although the two proposed control strategies achieve similar flow, their configurations differ in terms of time--space diagrams and velocity distributions. The GC strategy effectively maintains traffic flow while exhibiting a wide velocity distribution among vehicles. Conversely, the FC strategy maintains the width of the velocity distribution until reaching the metastable state; this implies that the FC strategy may lead to lower fuel consumption compared with the GC strategy.
The characteristics of each control strategy are evident in the fundamental diagrams and remain consistent up to 80\% penetration rate.

In conclusion, the CSOV model proposed in this study shows promising potential and warrants further exploration with additional control strategies. An important question for future research is to determine the driving behavior parameters and compare them with empirical experiments. For instance, studies such as \cite{malibari2022} utilize measured vehicle trajectory data from highways to assign values to a mathematical model. This approach could be applied to the CSOV model to anticipate the future traffic flow landscape in a scenario where autonomous vehicles become more prevalent. We can enhance the model's accuracy and applicability by incorporating real-world data; this would provide valuable insights for transportation planning and management.

\section*{Disclosure statement}
No potential conflict of interest was reported by the author(s).

\section*{Acknowledgements}
This research was partially supported by JSPS Grant-in-Aid for Scientific Research (C) JP17K05147, JP23K11139, and the graduate program Science and Technology for Global Leaders at Ochanomizu University\add{, and the Kansai University Fund for Supporting Young Scholars, 2022}.
We would like to thank Yuri Nakayama and Reika Matsunaga for useful discussions.

\bibliography{collection}

\end{document}